# Feasibility of Material Decomposition-based Photon-counting CT Thermometry

Nathan Wang[a]    Petteri Haverinen[b]

a. Department of Biomedical Engineering
Johns Hopkins University
Baltimore, Maryland, USA 21218
swang279@jhu.edu

b. Aalto Design Factory
Aalto University
Espoo, Helsinki, FI  02150
peheri.haverinen@gmail.com

## ABSTRACT

Over the past decades, thermal ablation procedures such as high intensity focused ultrasound (HIFU) have been developed vaporize cancerous tissues in a focal area. Thermal ablation has the potential to non-invasively eliminate tumors and treat other medical conditions with high efficacy and better patient outcomes. However, it is necessary for treatment to be coupled with real time temperature monitoring in order to deliver sufficient thermal dosage to the target while sparing surrounding, healthy tissues. To this end, computed tomography (CT) has been explored to offer fine spatial resolution at a rapid speed. However, current CT thermometry techniques are sensitive to heterogeneous tissue properties, imaging noise, and artifacts. To address this challenge, this paper utilizes the emerging photon-counting CT technology, performs tomographic thermometry based on material decomposition, and obtains excellent results in realistic numerical simulation. Specifically, three algorithms were designed to decouple material composition and thermal expansion from spectral CT reconstruction and compared in a comparative study. The best algorithm, referred to as the one-step algorithm, uniquely finds both material decomposition and the associated temperature at the same time in a closed form solution, which is robust to changes in tissue composition and generates temperature predictions with centigrade accuracy under realistic CT image noise levels.

## 1. INTRODUCTION

Non-invasive thermal ablation procedures, such as high intensity focused ultrasound (HIFU), heat a focal point up to several inches deep inside the human body [1]. In HIFU, local tissue temperatures can exceed 90 °C for several seconds, inducing necrosis in the targeted tumors. In the thermal ablation process, accurate and real-time temperature monitoring is needed to minimize damage to tissues surrounding the target. In practice, thermal probes are the most popular choice, which are invasive and restricted to a limited number of points. As a better solution, volumetric tomographic thermometry which would measure a three-dimensional heat map, which is optimal to guide thermal treatments and predict outcomes. To this end, three imaging modalities were adapted, including computed tomography (CT), magnetic resonance imaging (MRI), and ultrasound (US), which have their relative advantages and weaknesses. MRI thermometry is desirable for its rich tissue background, but it is expensive and slow [2]. US thermometry is portable and cheap but is subject to strong image artifacts and penetration limits [3]. CT is universally applicable for its high-resolution, fast speed, deep penetration, and cost-effectiveness. However, CT thermometry suffers from several drawbacks, especially the modeling complexity of heterogeneous tissue and measurement sensitivity to image noise and artifacts, as well as the need to minimize ionizing radiation dose [4], [5]. Given the recently emerging photon-counting CT, we realize that this technology can be developed to address many of the issues of CT thermometry simultaneously.

In general, heat introduced into a region of interest (ROI) causes a thermal expansion, which reduces the attenuation coefficient of the targeted tissue. Hence, the corresponding pixel intensity in the reconstructed image, measured in Hounsfield units (HU), is slightly decreased. This phenomenon is well known and can be linearly approximated as:

$$\Delta HU(T) \approx -[1000 + HU(T_0)]\alpha\Delta T \quad \{1\}$$

where α, the thermal expansion coefficient, is a tissue property [6]. Furthermore, experiments have shown a strong linear correlation between HU and temperature for water, bone, fat, and other tissues over the clinically relevant range of ~27 ℃ up to ~90 ℃ [4]. Thus, to enable 2D or 3D temperature imaging, a CT image could be segmented into several component tissues where their HU values could be converted to temperature via the tissue-appropriate linear regression models [7]. However, this method does not account for significant deviations in tissue properties as a result of inter-patient genetic variability or heat-induced necrosis from thermal ablation.

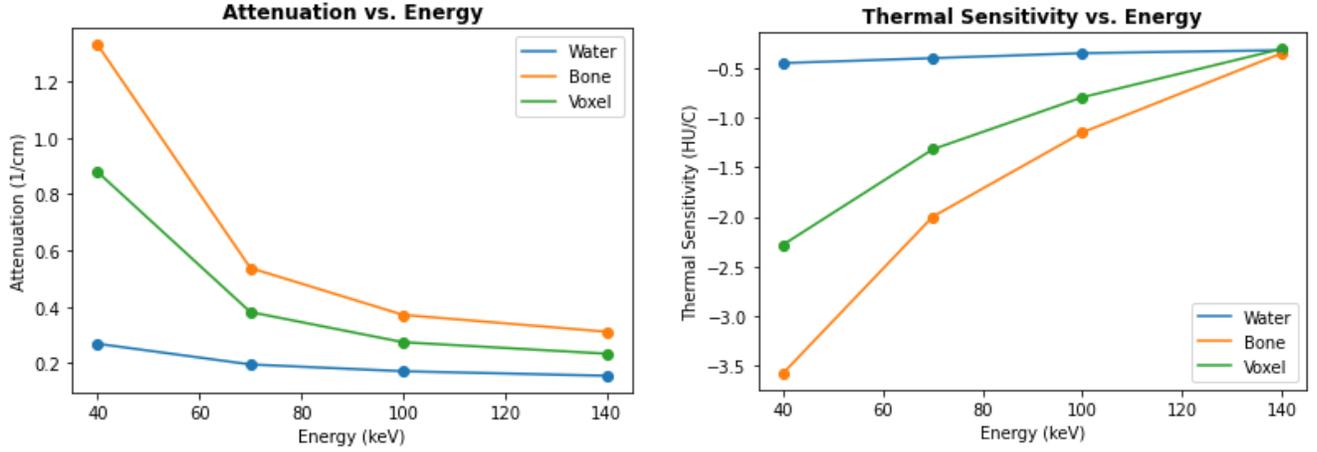

*Figure 1: (a). Attenuation vs. energy trend for water, bone, and a simulated voxel. (b). Thermal sensitivity vs. energy trend for water, bone, and a simulated voxel.*

Recently, the development of photon-counting CT, which resolves a polychromatic X-ray source into several energy bins, is recognized as a key breakthrough in the medical imaging field [6]. In addition to being less prone to beam hardening artifacts, spectral CT characterizes a tissue by its volume or mass fraction of selected component materials through material decomposition, which has a major implication for CT thermometry. Here we present and compare three algorithms for photon-counting CT thermometry based on material decomposition and thermal expansion. The goal of this study is to show that these approaches yield accurate and reliable temperature predictions across material compositions and noise levels, and under realistic conditions in simulation.

In the next section, we begin by outlining the theoretical background and assumptions behind the numerical simulations as well as the derivation of each of the three temperature prediction models. In the third section, we analyze the performance of each method over different experimental conditions. In the last section, relevant questions and issues are discussed and promising directions of related research are identified.

## 2. METHODS AND MATERIALS

The simulation was performed in our in-house simulation environment code in Python, with energy-dependent attenuation data obtained from NIST and thermal sensitivity values gathered from literature [4], [8]. Water and cortical bone are selected as the bases for material decomposition. Without loss of generality, the spectral CT is assumed to have four energy channels. The first method treats material decomposition and temperature prediction as sequential steps. The second method extends the first by refining the predictions with gradient optimization. The third method simultaneously obtains the volume fractions of base materials and temperature as the solution to a matrix system.

### 2.1 Simulation Conditions

It is known that the linear attenuation coefficient of an arbitrary material can be written as the sum of the attenuation coefficients of two basis materials, provided that they are sufficiently different. The relationship can be formulated as the following linear combination:

$$\mu(E,T) = V_w\mu_w(E,T) + V_b\mu_b(E,T) \quad \{2\}$$

where E is energy, T is temperature, V is a volume fraction, and the subscripts 'w' and 'b' correspond to 'water' and 'bone' as the low and high attenuation bases respectively. An analogous relationship in terms of HU can be obtained by subtracting $\mu_w(-V_w - V_b)$ from both sides, noting that

$-V_w - V_b = -1$, multiplying both sides by $\frac{\mu_w}{1000}$, and rearranging to arrive at:

$$\Delta HU(T) = V_w L_w(T) + V_b L_b(T) \quad \{3\}$$

where $L_w$ and $L_b$ are regression models in the form of $L(T) = \alpha T + \beta$ that map from T in degrees Celsius to L(T) in HU. The thermal sensitivity α has units of HU/°C and β is an offset term with units of HU. This approximation is made based on prior studies showing a strong linear correlation between increasing temperature and decreasing HU. Note that the energy variable E is omitted because the energy bins for the photon-counting detectors are assumed fixed. In this study, the energy bins are centered at 40 keV, 70 keV, 100 keV, and 140 keV respectively.

We can substitute and re-arrange the terms in the above equation by introducing a weighted slope $\alpha_{w,b} = V_w \alpha_w + V_b \alpha_b$ and similarly, a weighted offset $\beta_{w,b} = V_w \beta_w + V_b \beta_b$, and constructing an "aggregate" regression model:

$$HU(T) = \alpha_{w,b} T + \beta_{w,b} \quad \{4\}$$

The intercept term can be eliminated if we consider only the change in HU:

$$\Delta HU(\Delta T) = \alpha_{w,b} \Delta T \quad \{5\}$$

with respect to some reference temperature such as $T_0 = 23$ °C for this study. Observe that given the formula for weighted slope and Equation 2, the energy-dependent attenuations and thermal sensitivity properties can be generated for any material, defined by its volume fraction decomposition, provided that the attenuation and thermal sensitivity properties of the two basis materials are known (Fig. 1). The energy-dependent attenuation values for water and cortical bone were readily available from NIST while the thermal sensitivity values were researched from literature.

In the simulation, we use six phantoms with the following compositions: (0% water, 100% bone), (20% water, 80% bone), (40% water, 60% bone), (60% water, 40% bone), (80% water, 20% bone), (100% water, 0% bone). The attenuation values are computed by Equation 2. Each phantom is composed of a 10x10x1 region of voxels where uniform heating is applied from 23 °C (ΔT = 0°C) to 90 °C (ΔT = 67°C) with a step size of 1 °C. The change in attenuation as a result of heating is computed by Equation 5. Figure 2 displays an example of heating applied to a single voxel.

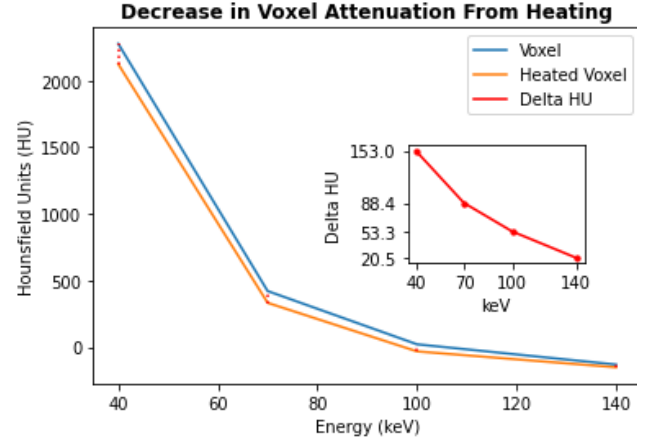

*Figure 2. Decrease in HU of 40% water 60% bone phantom heated to 90 °C*

Note that the HU value falls below zero at the higher energy extremes due to the decision to use 0.25 cm$^{-1}$, which is the approximate attenuation of water at 40 keV, as the normalization term in the conversion from linear attenuation to HU.

For every phantom at each temperature level, random normal noise is independently applied to the 100 voxels as the ROI. The standard deviation of the noise ranges from 0 to 15 HU with step size of 1 HU. The three temperature prediction algorithms are respectively applied under each condition to assess their performance and stability with respect to different material compositions and image noise levels.

**2.2 Method I: Sequential Temperature Prediction**

In this simulation, it is assumed that the spectral CT has four energy bins, resulting in four equations, plus the conservation of mass condition, in terms of three unknown volume fractions:

$$\begin{cases} \mu(E_1) = V_w \mu_w(E_1) + V_b \mu_b(E_1) \\ \vdots \\ \mu(E_4) = V_w \mu_w(E_4) + V_b \mu_b(E_4) \\ 1 = V_w + V_b + V_a \end{cases} \quad \{6\}$$

where $E_1$, $E_2$, $E_3$, and $E_4$ correspond to increasing energy of the photon-counting detector. $V_a$ is the volume fraction of air in a sample, which acts as a slack variable and has negligible contribution to attenuation. In matrix form, this is equivalent to solving the following system [9]:

$$\begin{bmatrix} \mu(E_1) \\ \vdots \\ \mu(E_4) \\ 1 \end{bmatrix} = \begin{bmatrix} \mu_w(E_1) & \mu_b(E_1) & 0 \\ \vdots & \vdots & 0 \\ \mu_w(E_4) & \mu_w(E_4) & 0 \\ 1 & 1 & 1 \end{bmatrix} \begin{bmatrix} V_w \\ V_b \\ V_a \end{bmatrix} \quad \{7\}$$

The solution may be interpreted to mean that an even mixture of water, bone, and air at the computed proportions would exhibit equivalent attenuation properties as the original sample. Note that the 5x3 system is over-constrained, so extraneous equations can be excluded, or the Moore-Penrose pseudoinverse can be employed. Once the volume fractions are obtained, back solving Equation 5 readily gives us $T_{pred} = \frac{\Delta HU}{\alpha_{w,b}} + 23°C$. Clearly, each of the four energy channels report distinct ΔHU values. Additionally, a unique $\alpha_{w,b}$ needs to be computed with respect to each channel. Hence, four temperature predictions are averaged together as the final prediction.

When a sample is heated, the terms $\mu_w(E, T)$ and $\mu_b(E, T)$ in Equation 2 should be updated to reflect the change in temperature. However, method I assumes that the temperature is constant at 23°C when performing material decomposition. This is because the attenuation data available from NIST was recorded at that reference temperature. Hence, for temperatures greater than 23°C, there is an error in predicted volume fractions, even if the attenuation values are measured with zero noise.

### 2.3 Method II: Iterative Temperature Prediction

To method II, a first pass prediction of the volume fractions and temperature is performed with method I. The motivation here is to perform gradient descent to make small but significant adjustments to the initial predictions. To do so, Equation 2 and Equation 5 can be combined and expressed as a loss function:

$$J = \sum_{i=1}^{4} [\mu(E_i, \Delta T) - V_w \mu_w(E_i) - V_b \mu_b(E_i) - \Delta T(\alpha_w(E_i) + \alpha_b(E_i))]^2 \quad \{8\}$$

Recall that $\alpha_{w,b} = V_w \alpha_w + V_b \alpha_b$. Additionally, assuming the volume fraction of air is negligible, the penalty term $(1 - V_w - V_b)^2$ is added to the cost. At the unique true values of $V_w$, $V_b$, and $\Delta T$, the objective function is zero and the gradient is the zero vector. The partial derivatives of the objective function are as follows:

$$\begin{cases} \frac{\partial J}{\partial V_w} = \left(-2 \sum_{i=1}^{4} \mu_w(E_i)\right) \sqrt{J(V_w, V_b, \Delta T)} \\ \frac{\partial J}{\partial V_b} = \left(-2 \sum_{i=1}^{4} \mu_b(E_i)\right) \sqrt{J(V_w, V_b, \Delta T)} \\ \frac{\partial J}{\partial (\Delta T)} = \left(-4 \sum_{i=1}^{4} [\alpha_w(E_i) + \alpha_b(E_i)]\right) \sqrt{J(V_w, V_b, \Delta T)} \end{cases} \quad \{9\}$$

Adjusting each variable in the opposite direction of the gradient reduces the objective functions step by step toward the optimal solution:

$$\theta_J := \theta_j - \alpha \frac{\partial J(V_w, V_b, \Delta T)}{\partial \theta_j} \quad \{10\}$$

where $\theta_J$ represents a parameter, which in this case is either $V_w$, $V_b$, or $\Delta T$. α is the learning rate empirically set to 0.01. The gradient term is also normalized such that the greatest change in volume fraction and temperature in one step is 0.01% and 1 °C respectively. This is done to ensure a stable trajectory to the optima.

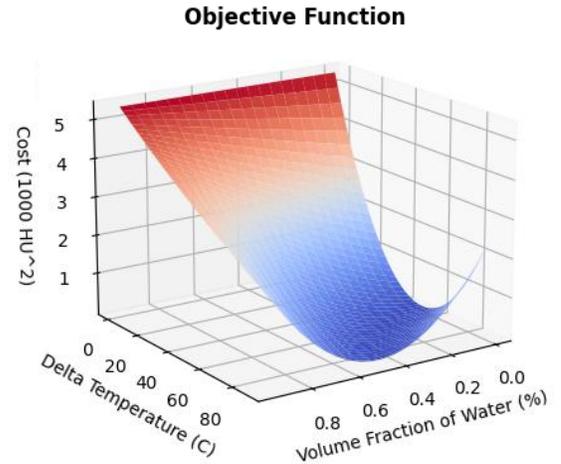

*Figure 3. Gradient descent objective function with ground truth at 40% water, 60% bone, and 70 °C*

### 2.4 Method III: Simultaneous Temperature Prediction

To compensate for the effect of thermal expansion on attenuation properties, corrections to Equation 2 can be made in the follow manner:

$$\mu(E, \Delta T) = V_w[\mu_w(E) - \alpha_w(E)\Delta T] + V_b[\mu_b(E) - \alpha_b(E)\Delta T] \quad \{11\}$$

This relationship can be converted into vector form using the fact that the spectral CT discretizes the X-ray sources into four discrete energy bins.

$$\begin{bmatrix} \mu(E_1) \\ \mu(E_2) \\ \vdots \\ \mu(E_4) \end{bmatrix} = V_w \begin{bmatrix} \mu_w(E_1) \\ \mu_w(E_2) \\ \vdots \\ \mu_w(E_4) \end{bmatrix} - \Delta T V_w \begin{bmatrix} \alpha_w(E_1) \\ \alpha_w(E_2) \\ \vdots \\ \alpha_w(E_4) \end{bmatrix} + V_b \begin{bmatrix} \mu_b(E_1) \\ \mu_b(E_2) \\ \vdots \\ \mu_b(E_4) \end{bmatrix} - \Delta T V_b \begin{bmatrix} \alpha_b(E_1) \\ \alpha_b(E_2) \\ \vdots \\ \alpha_b(E_4) \end{bmatrix} \quad \{12\}$$

Now, by defining $\gamma_1 = -\Delta T V_w$ and $\gamma_2 = -\Delta T V_b$, we can formulate the scalar coefficients in the above equation as the solution to a linear system where the left-hand side vector in Equation 7 is now set equal to:

$$\begin{bmatrix} \mu_w(E_1) & \mu_b(E_1) & 0 & \alpha_w(E_1) & \alpha_b(E_1) \\ \vdots & \vdots & 0 & \vdots & \vdots \\ \mu_w(E_4) & \mu_w(E_4) & 0 & \alpha_w(E_4) & \alpha_b(E_4) \\ 1 & 1 & 1 & 0 & 0 \end{bmatrix} \begin{bmatrix} V_w \\ V_b \\ V_a \\ \gamma_1 \\ \gamma_2 \end{bmatrix} \quad \{13\}$$

Once the solution vector is obtained via linear inversion, we have $\Delta T = \frac{-\gamma_1}{V_w}$ or $\frac{-\gamma_2}{V_b}$. If the volume fraction of water is very small, then the volume fraction of bone can be used to predict temperature and vice versa. If both volume fractions are sufficiently greater than zero, the two temperature predictions may be combined by averaging.

## 3. RESULTS

The performance of each of the three methods are analyzed under three criteria: Sensitivity to changes in material composition, robustness to varying levels of noise, and real-time performance. A summary of each method with respect to these metrics are reported in Table 1:

| Method | Composition Robustness | Noise Robustness | Run Time |
|---|---|---|---|
| One Step | ~ 0°C/1% | ~ 0.02 °C/HU | 13.7 ms/ROI |
| Two Step | ~ 0°C/1% | ~ 0.07 °C/HU | 18.7 ms/ROI |
| Iterative | ~ 0°C/1% | ~ 0.1 °C/HU | 50.9 ms/ROI |

*Table 1. Summary of the performance of each algorithm*

The specific data used to compile this summary is displayed in Figure 5 on the following page where the blue line "Two Step" refers to the sequential algorithm (2.2), the orange line "One Step" refers to the simultaneous temperature prediction (2.3), and the green line "Iterative" refers to gradient search method. The light shaded region extends to the 68% confidence interval around each data point and the solid-colored line indicates the general trend.

The composition robustness is reported in units of °C/1%, which measures the change in error corresponding to a 1% increase in the volume fraction of water. The performance of all three methods was not significantly affected by the material composition. Noise robustness is reported as the slope of the lines in Figure 5 and has units of °C/HU. Finally, run time is expressed in ms/ROI where the ROI is our 10x10 region of pixels. For example, if a CT scan has dimensions of 512x512, it possesses approximately 2,600 ROI's, leading to an overall computational time of 36 seconds using the "one step" method. Note that the simulation was performed on an AMD Ryzen 7 3700X CPU. With regard to matrix computations, performing the simulation on a dedicated GPU could potentially reduce the compute time by an order of magnitude [10]. Overall, it appears that the "one step" or "simultaneous" method possesses the most favorable qualities in each metric.

## 4. DISCUSSION AND CONCLUSION

CT thermometry based on photon-counting CT and material decomposition has significant promise to solve long standing challenges in the field, including the detrimental effects of noise and material heterogeneity on thermal predictions. The equations derived in this paper are theoretically true and stable in simulation. Particularly, the "one step algorithm" is a closed form solution that simultaneously computes the volume fractions and associated temperature with high accuracy and high speed. These results provide motivation for further experimental and clinical studies. Figure 4 depicts a CAD model of a phantom device designed by the authors for an experimental study.

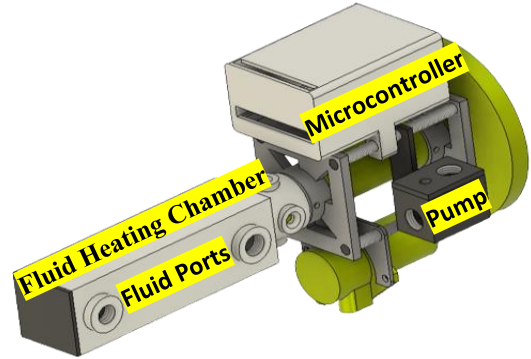

*Figure 4. CAD model of experimental phantom for uniform heating of homogeneous solutions*

With this device, fluid is pumped through the inlet and outlet ports into a rectangular chamber, where heating is controlled using a PID loop involving a thermometer, three 40-watt heating elements, and an Arduino microcontroller. The geometry of the heating chamber ensures all X-rays pass through approximately equivalent profiles, allowing measurements to be averaged. First, water and a bone powder solution will be heated to establish the attenuation and thermal sensitivity properties of the base materials. Then, different mixtures will be heated in order to compare the CT predicted temperature value with the gold standard thermometer reading.

For imaging a three-dimensional thermal distribution, a plastic sphere can be placed into a uniform heat bath for a set amount of time. Using existing finite element analysis methods, the heat distribution inside the sphere can be estimated [11] and compared to the heat distribution predicted using CT thermometry.

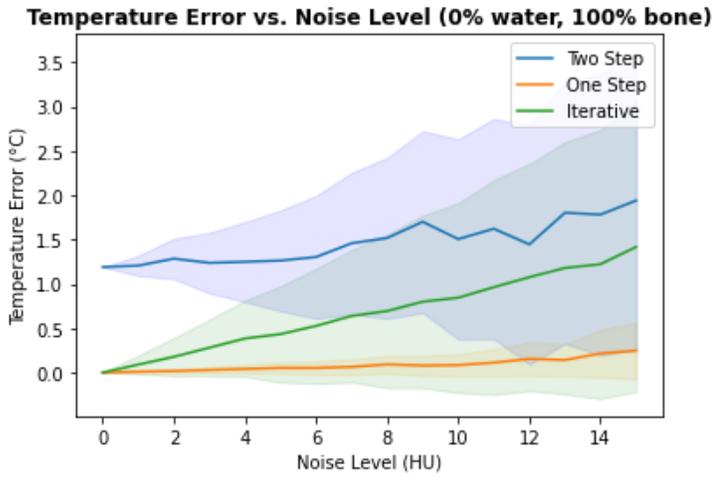
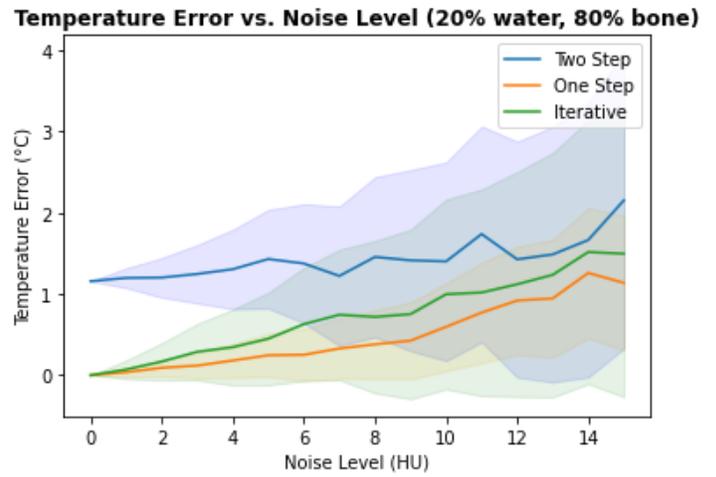
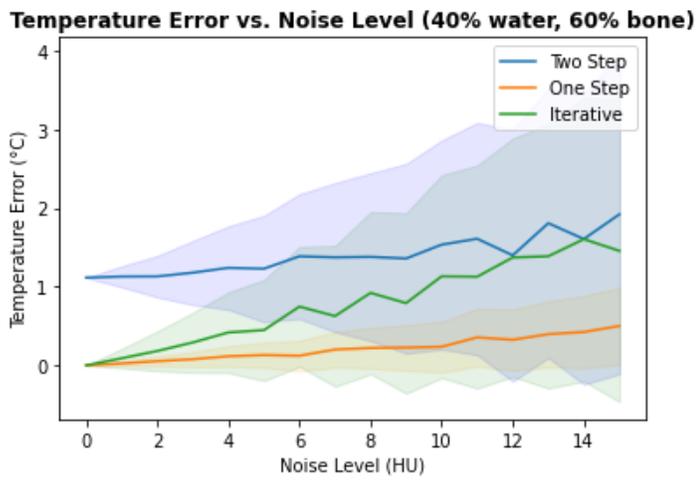
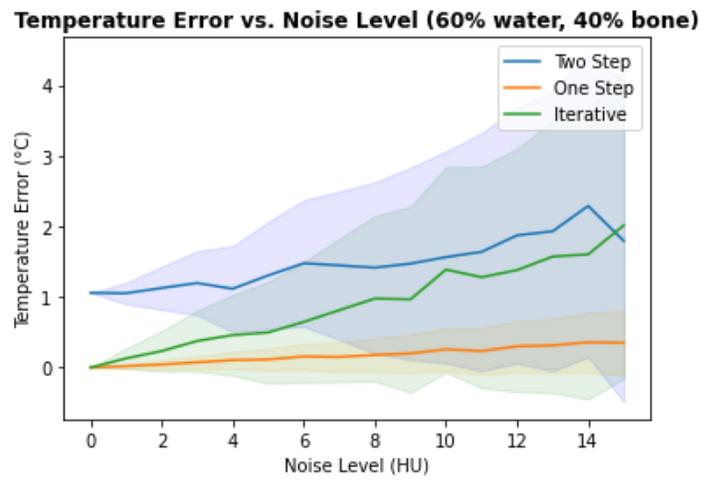
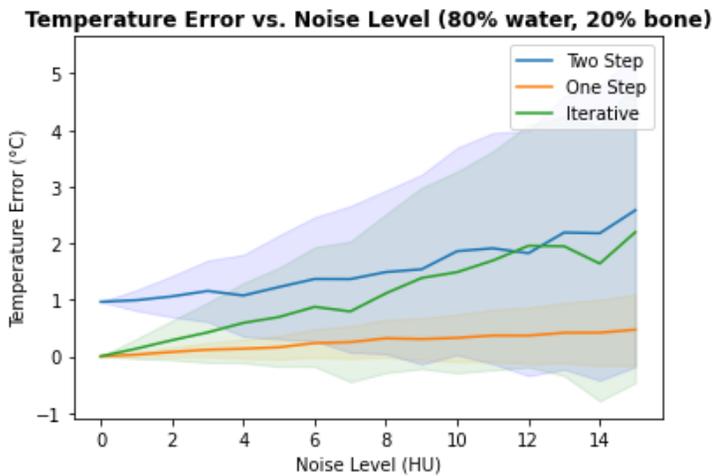
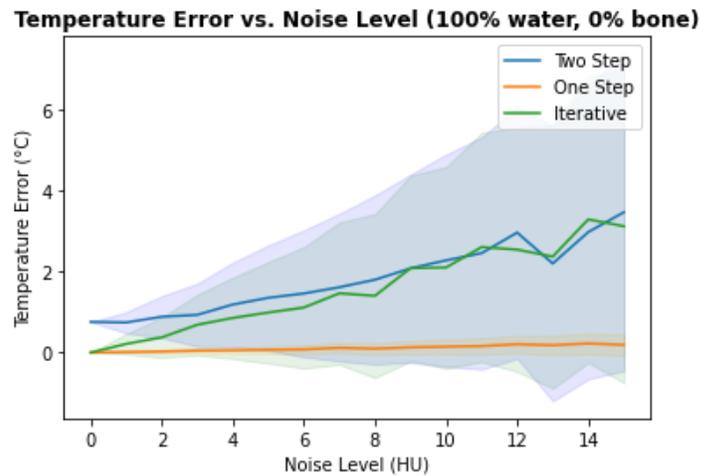

*Figure 5. Grid of results for the performance of the sequential/two step (blue), simultaneous/one step (orange), and gradient descent/iterative (green) over increase noise levels. The general trend is plotted with a dark colored line while the 68% confidence interval is shaded in with the light color.*

The applications of synergistic technologies should also be considered. In regard to radiation dose, deep learning reconstruction methods which can generate high quality clinical images from sparse/few-view data can significantly reduce a patient's exposure to radiation [12]. Additionally, interior tomography can be applied to image only the specific ROI where thermal ablation is being performed. The narrower beam profile reduces radiation dose and also increases contrast resolution [13].

As photon-counting CT become the gold standard in medical imaging, a further advantage of these methods is that the heat distribution can be obtained at no extra cost. Thus, any spectral CT dataset can be converted into heat maps, providing an entirely new domain of information that may better inform disease classification, localization, and prognostication. For example, it would be interesting to modify current deep neural networks for diagnosis of tuberculosis to include heat map information. Prior research has suggested that the difference in temperature between inflamed and healthy tissue can vary by several degrees Celsius [14]. Furthermore, these local changes in temperature were found to originate from local biochemical activity. Thus, heat maps may provide high-level insight into the cellular activity of a patient.

## 5. REFERENCES

[1] Z. Izadifar, Z. Izadifar, D. Chapman, and P. Babyn, "An Introduction to High Intensity Focused Ultrasound: Systematic Review on Principles, Devices, and Clinical Applications," *J. Clin. Med.*, vol. 9, no. 2, p. 460, Feb. 2020, doi: 10.3390/jcm9020460.

[2] V. Rieke and K. B. Pauly, "MR Thermometry," *J. Magn. Reson. Imaging JMRI*, vol. 27, no. 2, pp. 376–390, Feb. 2008, doi: 10.1002/jmri.21265.

[3] E. S. Ebbini, C. Simon, and D. Liu, "Real-time Ultrasound Thermography and Thermometry," *IEEE Signal Process. Mag.*, vol. 35, no. 2, pp. 166–174, Mar. 2018, doi: 10.1109/MSP.2017.2773338.

[4] F. Fani, E. Schena, P. Saccomandi, and S. Silvestri, "CT-based thermometry: An overview," *Int. J. Hyperthermia*, vol. 30, no. 4, pp. 219–227, Jun. 2014, doi: 10.3109/02656736.2014.922221.

[5] A. Mahnken and P. Bruners, "CT thermometry: Will it ever become ready for use?," *Int. J. Clin. Pract. Suppl.*, vol. 65, pp. 1–2, Apr. 2011, doi: 10.1111/j.1742-1241.2011.02651.x.

[6] M. Ahmed, C. L. Brace, F. T. Lee, and S. N. Goldberg, "Principles of and advances in percutaneous ablation," *Radiology*, vol. 258, no. 2, pp. 351–369, Feb. 2011, doi: 10.1148/radiol.10081634.

[7] N. Wang, M. Li, H. Shan, and P. Yan, "Deep learning based CT thermometry for thermal tumor ablation," in *Developments in X-Ray Tomography XII*, San Diego, United States, Sep. 2019, p. 62. doi: 10.1117/12.2530771.

[8] "NIST: X-Ray Mass Attenuation Coefficients - Table 4." https://physics.nist.gov/PhysRefData/XrayMassCoef/tab4.html (accessed Feb. 27, 2022).

[9] W. Cong, Y. Xi, P. Fitzgerald, B. De Man, and G. Wang, "Virtual Monoenergetic CT Imaging via Deep Learning," *Patterns*, vol. 1, no. 8, p. 100128, Oct. 2020, doi: 10.1016/j.patter.2020.100128.

[10] University of Ljubljana, Faculty of computer and information science, Slovenia., T. Dobravec, and P. Bulić, "Comparing CPU and GPU Implementations of a Simple Matrix Multiplication Algorithm," *Int. J. Comput. Electr. Eng.*, vol. 9, no. 2, pp. 430–438, 2017, doi: 10.17706/IJCEE.2017.9.2.430-438.

[11] H. Binous and B. Higgins, "Transient Cooling of a Sphere," *WOLFRAM Demonstrations Project*. http://demonstrations.wolfram.com/TransientCoolingOfASphere/ (accessed Feb. 27, 2022).

[12] C. Arndt, F. Güttler, A. Heinrich, F. Bürckenmeyer, I. Diamantis, and U. Teichgräber, "Deep Learning CT Image Reconstruction in Clinical Practice," *ROFO. Fortschr. Geb. Rontgenstr. Nuklearmed.*, vol. 193, no. 3, pp. 252–261, Mar. 2021, doi: 10.1055/a-1248-2556.

[13] G. Wang and H. Yu, "Meaning of Interior Tomography," *Phys. Med. Biol.*, vol. 58, no. 16, pp. R161–R186, Aug. 2013, doi: 10.1088/0031-9155/58/16/R161.

[14] M. Segàle, "THE TEMPERATURE OF ACUTELY INFLAMED PERIPHERAL TISSUE," *J. Exp. Med.*, vol. 29, no. 3, pp. 235–249, Feb. 1919.